\documentclass[
]{ceurart}

\usepackage{listings}
\begin{document}

\copyrightyear{2023}
\copyrightclause{Copyright for this paper by its authors.
  Use permitted under Creative Commons License Attribution 4.0
  International (CC BY 4.0).}

\conference{In: B. Combemale, G. Mussbacher, S. Betz, A. Friday,
    I. Hadar, J. Sallou, I. Groher, H. Muccini, O. Le Meur,
    C. Herglotz, E. Eriksson, B. Penzenstadler, AK. Peters,
    C. C. Venters.  Joint Proceedings of ICT4S 2023 Doctoral
    Symposium, Demonstrations \& Posters Track and
    Workshops. Co-located with ICT4S 2023. Rennes, France, June 05-09,
    2023.}

\title{A technological framework for scalable ground-up formation of
  Circular Societies}


\author[1]{Anant Sujatanagarjuna}[%
orcid=0000-0003-1376-407X,
email=anant.sujatanagarjuna@tu-clausthal.de
]
\cormark[1]
\address[1]{Technical University of Clausthal, Institute for Software and Systems Engineering (ISSE) Arnold-Sommerfeld-Str. 1, 38678 Clausthal-Zellerfeld, Germany}

\cortext[1]{Corresponding author.}

\begin{abstract}
  The Circular Economy (CE) is regarded as a solution to the
  environmental crisis. However, mainstream CE measures skirt around
  challenging the ethos of ever-increasing economic growth,
  overlooking social impacts and under-representing solutions such as
  reducing overall consumption. Circular Societies (CS) address these
  concerns by challenging this ethos. They emphasize ground-up social
  reorganization, address over-consumption through sufficiency
  strategies, and highlight the need for considering the complex
  inter-dependencies between nature, society, and technology on local,
  regional and global levels. However, no blueprint exists for forming
  CSs. An initial objective of my thesis is exploring existing
  social-network ontologies and developing a broadly applicable model
  for CSs. Since ground-up social reorganization on local, regional,
  and global levels has compounding effects on network complexities, a
  technological framework digitizing these inter-dependencies is
  necessary. Finally, adhering to CS principles of transparency and
  democratization, a system of trust is necessary to achieve
  collaborative consensus of the network state.
\end{abstract}


\maketitle


\section{Foreword}

I would like to thank the organizers of the ICT4S Doctoral Symposium
for creating this platform, and allowing the participation of
prospective doctoral researchers like myself. The following text
details the doctoral thesis topic that I have built for myself while
doing research in the \href{https://www.etce-lab.com}{Emerging
  Technologies for the Circular Economy} research group, under the
primary supervision of Dr. Benjamin Leiding. By participating in the
symposium, I hope to gain valuable feedback from interactions with
other researchers from diverse backgrounds, which would help me to
solidify this research direction that I have built as my doctoral
dissertation topic.

\pagebreak
\section{Background and Motivation}

The abundance of cheaply exploitable natural resources and human labor
has fueled a global system of production and consumption, which
follows a uni-directional pattern of "take, make and dispose";
beginning with material and resource extraction, followed by applying
(often non-renewably sourced) energy and human labor, in order to
manufacture products that are subsequently sold to consumers, who
discard them when they are of no further use~\cite{ellen}.  This
economic model, popularly known as the Linear Economy (LE), has been
the driving force of anthropogenic climate-change and its associated
widespread socio-ecological damage.

\subsection{Circular Economy}

While in the LE, materials are extracted from nature and are used to
manufacture products, only to be eventually turned into waste; the
model of a Circular Economy (CE) employs strategies such as sharing,
leasing, reusing, re-manufacturing and recycling to keep existing
materials and manufactured products in use for as long as possible.
The CE aims to design a system of production and consumption that
eliminates the concept of "waste", by designing products that are
optimized for cycles of disassembly and reuse~\cite{ellen}. The goal
of this redesign, is to close the loop in the system of "take, make
and dispose"; thus reducing material and resource extraction from
nature, the creation of waste, and environmental pollution.

In pursuit of similar goals, several governing bodies have adopted
this model in recent years. For instance, the EU Commission has
drafted a circular economy action plan as one of the main building
blocks of the European Green Deal~\cite{eu}.

While the CE is a necessary and important step towards transitioning
away from the destructive LE, CE in practice has been cricised in
recent research as being insufficient to achieve the transformational
change necessary to address the current socio-ecological crisis.

\subsection{Criticisms of the Circular Economy}

The central cause of the current socio-ecological crisis lies in the
unidirectional logic that characterizes global systems of consumption
and production today. This logic is fueled by industrialization and
the narrative of free-market capitalism, whose primary function is to
maximize economic value of natural and human resources that are
converted into market commodities~\cite{sixprop}. CE measures in
practice however, tend to focus on recovery rates, resource
efficiency, and waste reduction while overlooking aspects that can
challenge this fundamental logic. For instance, while the necessity
for radical transformation of the systems of production and
consumption has been acknowledged by proponents and stakeholders of CE
measures alike, the positive environmental potential of the concept of
sufficiency is paradoxically disregarded for being too radical, and is
excluded from current CE debates~\cite{sufficiency}. The concept of
sufficiency is tightly linked with sustainability. If human society
aims to be sustainable, i.e., satisfying the needs of the present
without compromising the needs of future generations, it must also aim
to consume enough to ensure universal human social well-being and
quality of life, while restricting this consumption within the
confines of the Earth's biocapacity~\cite{suffstudy}. In other words,
we must also practice sufficiency.

Another criticism of current CE measures, is that they often also
display a lack of consideration towards social sustainability
alongside economic and ecological sustainability~\cite{social}.

While the CE is a necessary step towards addressing the current
climate crisis, for a successful transition to a sustainable circular
economy that is truly within planetary boundaries, there is also the
need for an expansion of unidimensional value definitions to
multidimensional and holistic constructs that highlight the importance
of sufficiency and reducing resource consumption. The concept of a
Circular Society (CS) is such a holistic construct.

\subsection{Circular Society}
\label{sec:cs}

The CS is a holistic version of the CE, in which transition "requires
a fundamental reorientation and reorganization of practices and
processes in all areas of life --- from nutrition mobility and energy
use, to work models and housing concepts. This holistic vision,
hearkens back to the roots of the CE; "1. to adopt a system
perspective that considers the complex ways in which nature, society
and technology are interdependently interacting on a local, regional
and global level; 2. to aim for closed loops and organize production
and consumption practices in circular flows that imitates the
"eco-logic" of ecological systems; 3. to create a resilient
production and consumption metabolism taking the need for regeneration
of natural capital into account"~\cite{sixprop}. Unlike modern CE measures, which
implemented these concepts only through new business models and
technologies, the proponents of CS also call for a "re-valuation of
human labor and an enhanced role and conditions for productive work,
service provision and do-it-yourself (DIY) activities"~\cite{cs2}.

Aiming to challenge and transform the aforementioned unidimensional
value definitions, CS focuses on the creation of multidimensional
concepts of value creation "that define qualitative and quantitative
indicators for social and ecological value creation and which take
into account the many forms of work (care work, informal work,
community work, DIY) that contribute to societal
well-being"~\cite{cs2}.

The CS also "aims to establish a participatory, communitarian,
solidary and circular consumption and production system"~\cite{sixprop}, by
championing the idea of people as 'embedded' in these complex systems,
rather than as passive recipients~\cite{cs2}.

Jaeger-Erben et. al~\cite{cs2} formulated some central topics for a
"roadmap towards a CS", in which they highlight that unlike modern CE
measures, CS concepts must also emphasize the concept of sufficiency
through strategies such as "refuse, rethink and reduce". This focus on
sufficiency can shift the focus to alternative value definitions that
can aim to reduce, if not eliminate over-consumption.

The authors~\cite{cs2} also identify some core prerequisites that need to be
fulfilled, in order to foster community, collaboration and solidary
practices of a CS. Firstly, \emph{Accessibility and Transparency} are
recognized as central prerequisites for participation in the social
and economic practices of a CS, meaning access to natural resources as
well as education, health services and knowledge of the consumption
and production processes is shared rather than monopolized, and
"political and economic action is subject to the duty of
transparency". Secondly, "\emph{Democratization and Empowerment}
should create unconditional opportunities for participation and
engagement in political, economic and cultural
processes. Participation opportunities are linked with strategies for
activation, capability boosting, and empowerment." A focus on
encouraging sufficiency practices is also a key ingredient in
achieving a circular society.

\section{Research Gap}
\label{sec:rg}

There is a limit to what an individual alone can accomplish to
incorporate sufficiency practices in their life. For instance, as
individual computer scientists that recognize the socio-ecological
damage associated with the lifecycle of smartphones, we can refuse to
buy new smartphones, and rather utilize our expertise by repurposing
electronics that we already possess, and build our own software
alternatives to fit our daily needs. However, at least for me
personally, the expertise ends here; since I would be unable to repair
or build custom hardware components to sustain such a device
forever. At a certain point, the lack of expertise of a single
individual would be a barrier preventing them from pushing this
practice further. Thankfully, for this specific scenario, the concept
of a Hackerspace~\cite{hackerspace} is already quite popular. A
Hackerspace is a collaborative workspace for people to work on
projects while sharing tools and knowledge to realize
projects. Generalizing, in order to break such barriers, individuals
would need to collaborate and work together in order to practice
sufficiency strategies that they could not have achieved on their
own. Put together, several individuals can form a collaborative
environment that collectively finds creative solutions to
sustainability challenges; a circular society.

A similar analogy can be made on a slightly more macro scale:
individual local CS might also have similar limitations that they can
only overcome by networking with other CSs. Working inductively, there
needs to be perpetual scaling-up/networking of CSs, to actualize the
goal of a socially and environmentally sustainable global circular
society.

Jaeger-Erben et. al. ~\cite{sixprop} recognize that "fertile ground"
needs to exist for these innovative practices to be ubiquitous. Such a
fertile ground needs to be capable of networking individuals together
to facilitate these practices. There is however, no "blueprint" for
the creation and organization of CSs.

As mentioned in Section~\ref{sec:cs}, general guiding principles for
CSs exist. However, the idea of CSs is still relatively new, and there
are already several examples of the principles of the CS being put
into practice, e.g., the Free and Open Source Software (FOSS)
movement~\cite{foss}, solidarity economics~\cite{solidarityeconomics},
micro-energy cooperatives~\cite{energycoops}, eco-villages and
co-housing projects~\cite{ecovillages}. Exploring the commonalities
and differences between such societies, is a necessary step in
understanding how to replicate the fertile ground that facilitates
their operation.

Also, the perpetual scaling-up/networking of new and existing
societies towards the goal of a global CS would require a generic
framework that works on an individual, societal and macro-societal
level, and that can streamline the formation and continuous operation
of these CSs: the so-called fertile ground. Since this networking will
exponentially increase in complexity, tracking the various
interdependencies between individuals, society, nature and technology,
in an analog manner would be infeasible. Hence, this would require
this network of interdependencies to be digitized, calling for a
technological framework (at least) consisting of:

\begin{enumerate}
\item An ontology (consisting of individuals, society, nature,
  technology, and their various relations), that is broadly applicable
  to, and serves as a means of understanding of the dynamics of
  collaborative practices within and between CSs, inspired and
  compatible with existing concepts such as solidarity economics,
  eco-villages, collaborative development, etc.
\item A common communication network that allows transparent and
  democratized inter- and intra- networking of CSs by forming a
  \emph{decentralized knowledge graph}~\cite{kg} that adheres to, and
  is described by the ontology.
\item A system of trust that is able to achieve (decentralized)
  consensus of the state of this decentralized knowledge graph.
\end{enumerate}

These three building-blocks can satisfy the previously mentioned
prerequisites identified by Jaeger-Erben et. al~\cite{sixprop}, for a
fertile ground that can foster the creation of CSs. The
democratization in the formation of the \emph{decentralized knowledge
graph} would allow for universal engagement in the sufficiency
practices followed by the various actors in the network. The
\emph{decentralized knowledge graph} being transparent, is also
universally accessible, and allows knowledge to be extracted and
understood based on the defined ontology, empowering others with
blueprints of successful formations of CSs for independent
formation. Additionally, the documentation of sufficiency practices,
could also allow for their evaluation in terms of positive
environmental impact.

\section{Research Questions}
The main goal of my thesis is to fill the gap in the existing research
identified in Section~\ref{sec:rg}. To this end, I define the main
research question for my thesis as follows:

\textbf{RQ: How to create a technological framework to facilitate the
  formation and networking of circular societies?}

For a more focused exploration of the main research question, I have
further sub-divided the main research question into the following:

\begin{itemize}
\item \textbf{RQ1: How to create an ontology that is broadly
  applicable to circular societies?}

\item \textbf{RQ2: How to network circular societies in a transparent,
  democratized, and decentralized knowledge graph?}

\item \textbf{RQ3: How to create a system of trust to achieve
  consensus in the decentralized knowledge graph?}
\end{itemize}

These sub-questions are suitably phrased in order to collectively
answer the main research question.

In order to answer each of these questions, a preliminary research
direction of my thesis would be to do systematic literature review
into various subjects, including but not limited to: existing
ontologies of social networks and frameworks for their development
such as OWL (the Web Ontology Language)~\cite{owl}, decentralized
knowledge graphs, technologies supporting decentralized autonomous
organizations~\cite{dao} (including non-blockchain~\cite{blockchain}
related solutions), and decentralized consensus
mechanisms~\cite{consensus}.

Secondary research directions of my thesis will also include exploring
methods of evaluating sufficiency practices for their positive
environmental impact using this technological framework.

\section{Acknowledgments}

I would like to thank my members of my family, friends and colleagues;
whose feedback was invaluable in writing this paper.








\appendix
\bibliography{references}
\end{document}